# Evaluating Cybersecurity Risks of Cooperative Ramp Merging in Mixed Traffic Environments

Xuanpeng Zhao, Ahmed Abdo, Xishun Liao, *Student Member, IEEE,* Matthew J. Barth, *Fellow, IEEE,* and Guoyuan Wu, *Senior Member*

*Abstract*—Connected and Automated Vehicle (CAV) technology has the potential to greatly improve transportation mobility, safety, and energy efficiency. However, ubiquitous vehicular connectivity also opens up the door for cyber-attacks. In this study, we investigate cybersecurity risks of a representative cooperative traffic management application, i.e., highway on-ramp merging, in a mixed traffic environment. We develop threat models with two trajectory spoofing strategies on CAVs to create traffic congestion, and we also devise an attack-resilient strategy for system defense. Furthermore, we leverage VENTOS, a Veins extension simulator made for CAV applications, to evaluate cybersecurity risks of the attacks and performance of the proposed defense strategy. A comprehensive case study is conducted across different traffic congestion levels, penetration rates of CAVs, and attack ratios. As expected, the results show that the performance of mobility decreases up to 55.19% at the worst case when the attack ratio increases, as does safety and energy. With our proposed mitigation defense algorithm, the system's cyber-attack resiliency is greatly improved.

*Index Terms*─Cybersecurity, attack-resilience, connected and automated vehicles, cooperative ramp merging, mixed traffic

## INTRODUCTION

The ever-increasing number of vehicles on our roads has negative impacts on safety, traffic efficiency, and environment. In terms of safety, according to the data from World Health Organization, road traffic injuries caused approximately 1.35 million deaths worldwide in 2016 [1]. Moreover, a study by MIT in Year 2013 indicated annually 53,000 premature deaths due to health problems caused by vehicle emissions [2]. To address these problems, connected and automated Vehicle (CAV) technology has the potential to reduce traffic accidents, to enhance the quality of life, and to improve the efficiency of our transportation system [3]. CAVs can not only perceive the surrounding environment with on-board sensors such as cameras, radar, and LiDAR, but also communicate with each other or with roadside infrastructure via vehicle-to-vehicle (V2V) or vehicle-to-infrastructure (V2I) communications. This enables CAVs and other road users to perform specific operations efficiently and collaboratively.

As a representative scenario for highway driving, ramp merging has received significant attention over the years. Various traffic control strategies have been applied at the ramp merging area to regulate vehicular inflow rates to avoid mainline traffic breakdowns. A widely used ramp management strategy for legacy vehicles is ramp metering, which governs on-ramp vehicles' entry to the mainline by traffic lights [4]. However, ramp metering often introduces stop-and-go maneuvers [5][6], which significantly degrade overall system performance. To address this issue, many researchers have leveraged CAV technology and developed advanced ramp merging systems [7], which can coordinate CAVs into platoons or closely-spaced strings to maximize the throughput while smoothing speed trajectories [5][9] Such cooperative highway on-ramp merging systems are expected to significantly reduce traffic congestion by sharing information and implementing appropriate control measures.

It is clear that legacy vehicles and CAVs (with varying penetration rates, connectivity capabilities, and automation levels) will have to share the roads during a transition period, which may span several decades. Therefore, it is more realistic and valuable to develop an effective ramp merging strategy for mixed traffic and investigate its performance in terms of safety, efficiency, and environmental sustainability [10][11]. Many researchers have dug into this problem with different methods [12][13][14], but they fail to address potential cybersecurity risks in communication and perception. To ensure resilient operation and safety of all road users, potential exposure of cyber-attacks [15] should be considered for real-world implementation, and a cybersecurity-awareness defense strategy should be carefully designed.

In this paper, we reveal the potential cybersecurity risk of highway ramp merging strategy under mixed traffic and provide an effective defense solution. More specifically, three major contributions are listed below:

- Two spoofing strategies are designed, which cannot be detected through inconsistent speed and position information, for the cooperative highway on-ramp merging strategy
- A minimum mean square estimation (MMSE)-based defense algorithm is devised to detect the attacks and mitigate their negative impacts on the entire traffic.
- A comprehensive evaluation is conducted for the cybersecurity risks of proposed merging strategy with these two attacks and the performance of corresponding defense strategy.

The remainder of this paper is organized as follows: Section 2 discusses relevant background information. Section 3 illustrates the proposed cooperative highway on-ramp merging algorithm for mixed traffic, threat model, and defense strategy. Section 4 presents the simulation setup and evaluates the results of different simulation cases. Conclusions and future work are discussed in Section 5.

## BACKGROUND

In this section, we briefly review relevant studies on CAV-based cooperative highway on-ramp merging applications and potential cybersecurity risks in CAV applications. We also introduce the simulation environment for analyzing traffic impacts of cyberattacks.

## A. CAV-based Cooperative Ramp Merging

Wang et al. [6] proposed a cooperative merging system based on V2X communications and adopted a microscopic traffic simulator to evaluate its performance in terms of highway capacity. The basic idea is to use Road Side Unit (RSU)-equipped infrastructure to collect On-Board Unit (OBU)-equipped vehicles' information in the form of Basic Safety Messages (BSM) [8] via V2V and V2I communications. With this information, the system can provide speed guidance to the involved CAVs to improve merging efficiency. Zhao et al. [9] developed a hierarchical CAV-based ramp management system that can perform cooperative merging maneuvers for CAVs at the individual ramp level and regulate the inflow rates of multiple ramps simultaneously. Some of these existing studies assume full penetration rate of CAVs for effective ramp merging control. The feasibility and performance of their algorithms in mixed traffic still need to be verified. Chen et al. [16] proposed a hierarchical control approach consisting of a tactical layer and an operational layer to enable efficient and safe merging operations. Rios-Torres et al. [17] put forward a coordination strategy that allowed vehicles to merge without creating congestions under collision avoidance constraints, thus reducing both fuel consumption and travel time. Davis et al. [10] showed that traffic congestion might be significantly mitigated even with a 50% penetration rate of Adaptive Cruise Control-enabled vehicles. It is noted that most of these ramp merging related studies for mixed traffic do not consider potential cybersecurity issues.

## B. Cybersecurity for CAV-based Applications

Bhat et al. [18] presented an overview of commonly seen security risks associated with both automotive radar and dedicated short-range communication (DSRC) systems, such as jamming and spoofing. According to Dibaei et al. [19], jamming attacks include Denial-of-service (DOS) attacks and Distributed Denial-of-service attacks (DDOS), while spoofing attacks contain black-hole attacks, Sybil attacks, and replay attacks. In this paper, we focus on man-in-the-middle attacks. Chen et al. [20] investigated cybersecurity vulnerabilities in the Intelligent Traffic Signal System (I-SIG) application and concluded that current signal control algorithms are highly vulnerable to data spoofing attacks, even only one single attacked vehicle. They also assumed that the Intelligent Traffic Signal System (I-SIG) [21] had utilized the Security and Credential Management System (SCMS) [22]. This required every BSM to be signed by the sender's digital certificate to ensure the message integrity, before CAVs and infrastructure were allowed to participate in further communications. Thus, the receiver could verify the sender's identity by the signature. In this study, we also assume that SCMS has been deployed to ensure that all BSMs are authenticated. Cui et al. [23] developed an evaluation platform that simulated vehicles' sensor errors and communication delays to investigate the effect of cyber-attacks on CACC. They deployed heuristic cyber-attacks on the 3rd vehicle in the platoon by assigning constant errors on the GPS and radar information since the preset time step. Such attacks can be easily detected due to the resultant large position jumps, lane overtaking, and inconsistent speeds/positions. In this paper, we propose two attack strategies that are hardly detected by checking data inconsistencies. Wang et al. [24] incorporated the intelligent driver model (IDM) and communication scheme to numerically analyze cyber-attack effects on connected automated vehicular platoons, without comprehensive evaluation in a microscopic simulation environment. Hadded et al. [25] deployed three different attack methods on the proposed collective perception-based on-ramp merging control algorithm and measured their impacts. However, they did not provide an effective defense method to mitigate the potential attacks. Xu et al. [26] focused on the sensor perception aspect and aimed to reveal security risks of on-board sensors. They proposed two defense strategies against their well-designed attacks on ultrasonic sensors to improve the system resilience and validated them in both simulation and the real world. Giraldo et al [27] proposed a moving target defense strategy for multi-vehicle systems to mitigate the impacts caused by cyber-attacks. Liu et al. [28] presented an attack-resistant location estimation approach based on the mean square error (MSE) of distance difference between the declared distance and the distance determined by the received signal strength index (RSSI) [29] of the radio signal to tolerate malicious attacks. RSSI is an indicator of the signal quality and widely used to calculate the received signal power [30]. In wireless channel models, received power is inversely proportional to the distance between the sender and receiver. Motivated by this idea, we develop a secure and resilient defense strategy in this study to detect and filter out malicious attacks. Although RSSI localization has relatively lower accuracy than other ranging techniques due to multipath radio signals propagation [31], it is considered as a cost-effective method for rough estimation of position [32][33] and is suitable for our study purpose.

## C. Simulation Environment

In this study, we adopt and utilize the VEhicular NeTwork Open Simulator (VENTOS) [34] to perform the simulation. VENTOS is a Veins extension simulator for modeling vehicular traffic flows, collaborative driving, and interactions among CAVs or between CAVs and infrastructure equipped with DSRC. DSRC [35] is a wireless communications standard featured with reliable and low-latency data transmission. More specifically, VENTOS combines the capabilities of both vehicular traffic simulation from Simulation of Urban Mobility (SUMO) [36] and communication network simulation from Objective Modular Network Testbed in C++ (OMNET++) [37]. SUMO is a highly portable, microscopic, and continuous traffic simulator designed to handle large roadway networks, while OMNET++ is a C++ simulation library, which can simulate complex communication networks with high fidelity.

## METHODOLOGY

In this section, we present the mixed traffic cooperative highway on-ramp merging algorithm to be used in the simulation, followed by an elaborate description of threat models and defense strategies. The general assumptions in this study are presented below.

## A. Assumptions

1) All CAVs consistently send BSMs, including positions, speeds, and accelerations, via V2I communications to the

RSU-equipped infrastructure. Furthermore, they exactly follow the speed guidance recommended by the infrastructure.
2) In the network with a multi-lane mainline segment, only CAVs that are involved in merging maneuvers (i.e., those vehicles on the merging lanes) would be controlled by the proposed algorithm and are susceptible to attacks by the malicious actor.
3) We assume the attacker can only modify the content of BSMs but not manipulate the radio signal strength within the effective area, which is defined as the overlapped region of both RSU's and the attacker's communication coverage (given that RSU is within the attacker's coverage).
4) We also assume SCMS is deployed and cannot be exploited, which means that the attacker cannot falsify signatures of senders.

*B. Cooperative Highway On-Ramp Merging Algorithm in Mixed Traffic*

The proposed cooperative highway on-ramp merging system relies on V2I communications. When CAVs on the on-ramp and rightmost lane of the mainline enter the communication range of the RSU-equipped infrastructure in the merging area, they broadcast their state information via BSMs. Once receiving the involved CAVs' states, the RSU will perform the proposed ramp merging algorithm (see details below) to determine the merging sequence and longitudinal speed for each CAV and broadcast this information to enable cooperative maneuvers. The overall system architecture is illustrated in Figure 1. Firstly, we specify upstream roadway segments with respect to the merging area (within the RSU communication range) with two types of zones: control zones and buffer zones. In buffer zones, the RSU can sort out the incoming CAVs based on their distances to the merging point. In control zones, the RSU sends recommended speeds back to respective CAVs. The flow chart of the system is shown in Figure 2.

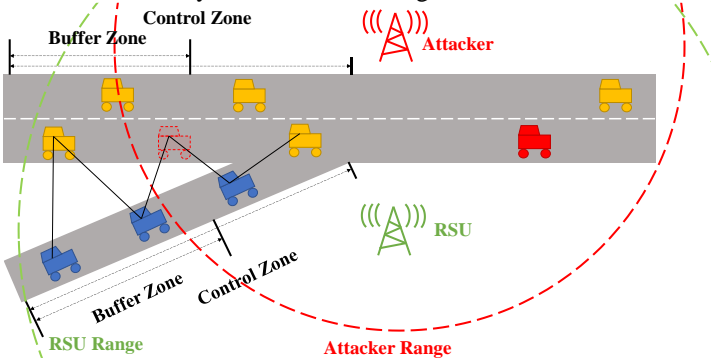

**Figure 1 Overall system architecture of the threat models.**

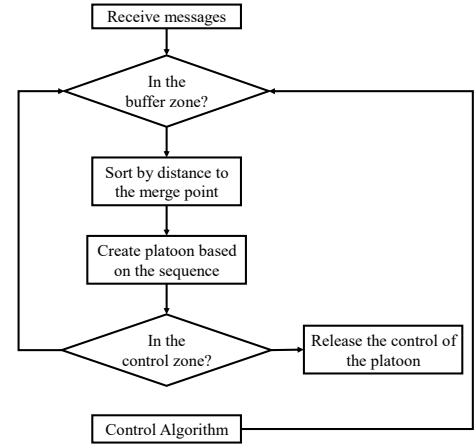

**Figure 2 System workflow of the cooperative ramp merging strategy for CAVs.**

1) Vehicle Sequence Algorithm

The RSU collects the information from CAVs on both the mainline and on-ramp via V2I communications and determines their entrance sequence based on their distances to the merging point. Moreover, if the distance between two CAVs is too far (e.g., due to legacy vehicles), we split the string into two and create a new leader for the new string. We then apply the customized motion control for each string of CAVs.

2) Motion Control Algorithm

Once CAVs are arranged into strings, the RSU will apply the following control algorithm to enable merging maneuvers in a cooperative manner. The control algorithm can compute the recommended speed for every following vehicle based on the state of its predecessor. Li et al. [38] provided a general car-following model that can be used to describe vehicles' longitudinal dynamics:

$$\dot{x}_n(t) = v_n(t) \qquad (1.)$$
$$\dot{v}_n(t) = f\left(s_{n(t)}, v_n(t), \Delta v_n(t)\right) \qquad (2.)$$

Based on this model, we conduct our control algorithm for the acceleration of ego vehicle:

$$a(t) = k_d * \left(S_{headway}(t) - S_{length}(t) - S_{safe\ gap}(t)\right) + \left(v_{front}(t) - v_{self}(t)\right) * k_v \qquad (3.)$$

where $a(t)$ is the acceleration of ego vehicle at time step $t$; $k_d$ and $k_v$ are control gains for distance and speed, respectively; $S_{headway}$ is the distance between ego vehicle and its predecessor; $S_{length}$ is the ego vehicle length; $S_{safe\ gap}$ is the safety distance gap which can guarantee the minimum clearance between two vehicles; $v_{front}(t)$ is the velocity of the front vehicle at time step $t$; $v_{self}(t)$ is the velocity of ego vehicle at time step $t$.

Thus, the recommended speed can be derived based on the acceleration.

$$v(t) = a(t) * t_{step} + v_{self}(t - 1) \qquad (4.)$$

where $t_{step}$ is the simulation time step; and $v_{self}(t-1)$ is the velocity of ego vehicle at time step $(t-1)$. The leader of each string is recommended to travel at the roadway speed limit. Based on our assumptions, CAVs that are not on the rightmost lane of mainline are not controlled by the merging algorithm. Similar to all legacy vehicles, their longitudinal behaviors are controlled by the Krauß car-following model [39][40], and their lateral maneuvers are governed by the LC2013 lane-changing model [41].

The Krauß model is defined below [40]:

$$v_{safe}(t) = v_l(t) + \frac{g(t) - g_{des}(t)}{\tau + \tau_b}, \quad (5.)$$

$$v_{des}(t) = \min\left[v_{\max, v(t)+a(v)\Delta t, v_{safe}(t)}\right], \quad (6.)$$

$$v(t + \Delta t) = \max[0, v_{des}(t) - \eta] \quad (7.)$$

$$x(t + \Delta t) = x(t) + v\Delta t \quad (8.)$$

where $g_{des}$ is the desired gap; $\tau$ is the reaction time of drivers; $\tau_b$ is the time for deceleration; $v_{safe}$ is the safe speed; $v_{des}$ is the desired speed; $v_{max}$ is the maximum speed; and $\eta$ is the random perturbation. Two major modifications on this model made by SUMO are: a) using the Euler-position update rule to make the safe speed formula suitable for maintaining safety; and b) using different deceleration capabilities to avoid violating safety [41].

The lane-changing model LC2013 was developed by J. Erdmann, which consists of four different motivations for lane-changing [41], such as *Strategic Change*, *Cooperative Change*, *Tactical Change*, and *Regulatory Change*. When a vehicle (either legacy vehicle or CAV) reaches the merging point, the lane-changing model will allow the vehicle to merge when it is safe. Otherwise, the vehicle will wait for a suitable gap to merge. To avoid the collision of a CAV with its preceding legacy vehicle in the mixed traffic simulation, we apply a heuristic gatekeeping logic by consistently comparing the recommended speed with the safe speed from the Krauß model, and choosing the lower one as the target speed, i.e.,

$$v_{target} = \begin{cases} v_{safe}, & v_{safe} < v_{recommended} \\ v_{recommended}, & v_{safe} > v_{recommended} \end{cases}$$

Please note that model predictive control (MPC) could also be an alternative to handle this type of safety constraints [16].

*C. Threat Models*

The proposed attack strategies aim at creating congestion while not being easily detected by simple defense approaches, such as inconsistency of vehicle's location and speed, teleporting, or same-lane-overtaking. Before the elaboration of attack strategies, we illustrate the scenario setup, as shown in Figure 1. The attacker is located near the RSU, and can intercept the BSMs broadcasted by equipped vehicles. Then, the attacker deploys man-in-the-middle attacks to modify the BSMs and resends them to the RSU. For example, the red vehicle is the attacked vehicle, and the dash-line icon represents the associated location where the attacker tries to spoof the RSU. In the following, we detail two non-trivial spoofing strategies.

1) **Emergency Stop spoofing**

When a target CAV enters the attacker's communication range, the attacker can consistently receive BSMs from this CAV. The attacker deploys man-in-the-middle attacks, which make the CAV's location information to be frozen at the respective entrance point and falsify its speed information to be zero, and then resend all this information to the RSU. In this case, the control algorithm will provide incorrect recommended speed to those following CAVs within the same string of the attacked vehicle to make them slow down or even completely stop. The pseudo-code is shown in Algorithm 1.

2) **Accumulative position drift spoofing**

In this type of attack, the attacker keeps receiving BSMs from the target CAV once it enters the attacker's communication range. Then, the attacker continuously generates falsified speed information over the time, i.e., speed profile that is slightly lower than the actual speed profile of the attacked vehicle. In addition, the spoofed location is computed based on the falsified speed. This slight inconsistency of location and speed may be considered as a result of signal loss or GPS errors. However, as the position drift accumulates, more severe impacts (e.g., congestion) on the upstream traffic would be shown up. The pseudo-code is shown in Algorithm 2.

| Algorithm 1: Emergency Stop Spoofing Algorithm |
|---|
| **INPUT:** BSM from target CAVs. |
| **INITIALIZE** Spoofing_flag = 0, Spoofing_Location |
| **IF** Received a BSM from a target CAV **THEN** |
|     Spoofing_BSM = BSM |
|     **IF** Spoofing_flag = 0 **THEN** |
|         Actual_Location = location contained in the BSM |
|         Spoofing_Location = Actual_Location |
|         Spoofing_flag = 1 |
|     **END IF** |
|     Spoofing_Speed = 0 |
|     Set Spoofing_Speed to be the speed in Spoofing_BSM |
|     Set Spoofing_Location to be the location in Spoofing_BSM |
|     Resend the Spoofing_BSM to the RSU |
| **END IF** |

| Algorithm 2: Accumulative Position Drift Spoofing Algorithm |
|---|
| **INPUT:** BSM from target CAVs. |
| **INITIALIZE** Spoofing_flag = 0, Spoofing_Location, Spoofing_Speed, Last_Time_Stamp, Accel = -2.5 |
| **IF** Received a BSM from a target CAV **THEN** |
|     Spoofing_BSM = BSM |
|     **IF** Spoofing_flag = 0 **THEN** |
|         Last_Time_Stamp = time stamp contained in the BSM |
|         Actual_Location = location contained in the BSM |
|         Actual_Speed = speed contained in the BSM |
|         Spoofing_Location = Actual_Location |
|         Spoofing_Speed = Actual_Speed |
|         Spoofing_flag = 1 |
|     **ELSE IF** Spoofing_Speed > 0 **THEN** |
|         Actual_Time_Stamp = time stamp contained in the BSM |
|         Delta_Time = Last_Time_Stamp - Actual_Time_Stamp |
|         Spoofing_Location = Spoofing_Location + Spoofing_Speed * Delta_Time + 0.5 * Accel * Delta_Time^2 |
|         Spoofing_Speed = Spoofing_Speed + Accel * Delta_Time |
|         Last_Time_Stamp = Actual_Time_Stamp |
|     **END IF** |
|     Set Spoofing_Speed to be the speed in Spoofing_BSM |
|     Set Spoofing_Location to be the location in Spoofing_BSM |
|     Resend the Spoofing_BSM to the RSU |
| **END IF** |

| Algorithm 3: | MSE Based Attack-Resilient Defense Algorithm |
|---|---|

```
INPUT:       A set of BSMs from CAVs: Current_Platoon.
OUTPUT:      Current_Platoon
INITIALIZE:  Flag = True, M = size of Current_Platoon, X_RSU,
Y_RSU, Threshold, MMSE = 99999
WHILE M > 1 AND Flag == True
    SUM_MSE = 0
    FOR i = 0 to M
        Dist = coarse distance measured by RSSI of Current_Platoon[i]
        X = x of location of Current_Platoon[i]
        Y = y of location of Current_Platoon[i]
        MSE = (Dist – sqrt((X – X_RSU)^2 + (Y – Y_RSU)^2))^2 / M
        SUM_MSE = SUM_MSE + MSE
    END FOR
    IF SUM_MSE < Threshold THEN
        BREAK
    ELSE
        SUM_MSE = 0
        FOR j = 0 to M
            FOR i = 0 to M
                IF j != i THEN
                    Dist = distance measured by RSSI of
                            Current_Platoon[i]
                    X = x of location of Current_Platoon[i]
                    Y = y of location of Current_Platoon[i]
                    MSE = (Dist – sqrt((X – X_RSU)^2 + (Y –
                            Y_RSU)^2))^2 / M
                    SUM_MSE = SUM_MSE + MSE
                END IF
            END FOR
            IF SUM_MSE < Threshold THEN
                Flag = False
                Remove Current_Platoon[j] from Current_Platoon
                BREAK
            ELSEIF SUM_MSE < MMSE
                MMSE = SUM_MSE
                Delete_Vehicle = Current_Platoon[j]
            END IF
        END FOR
        Remove Delete_Vehicle from Current_Platoon
        M = M – 1
    END IF
END WHILE
```

*D. Defense Strategy*

Motivated by Liu et al. [28], Motivated by Liu et al. [30], we define an MSE based attack-resilient defense strategy to detect and filter BSMs with spoofed data on the RSU side. We exploit the advantage that once RSU receives a BSM, the infrastructure can get the BSM's RSSI. The RSSI is approximately inversely proportional to the distance between the sender and the receiver. Our defense strategy aims to find the outlier(s) with larger error(s) compared to others. Therefore, it does not require accurate transmission distances that would be challenging for RSSI-based estimation. In this case, RSU receives BSMs from both CAVs and the attacker, and feeds them into the vehicle sequence algorithm to identify strings. For each string, we compute the mean square errors (MSEs) of distance measurements based on the statistics of RSSI and location information embedded in the received BSMs. Thus,

$$\varsigma^2 = \frac{\sum\left(\delta_i - \sqrt{(x_i - x)^2 + (y_i - y)^2}\right)^2}{m} \quad (9.)$$

where $\delta_i$ is the distance measured by RSSI for the $i$th CAV; $x_i$ and $y_i$ represent the latitude and longitude, respectively, contained in the BSM of the $i$th CAV; $x$ and $y$ are the GPS coordinates of RSU; and $m$ is the total number of CAVs in this string. We then define a threshold τ by the average MSE when there are no attacks:

$$\frac{\sum \varsigma_j^2}{n} < \tau^2 \quad (10.)$$

where $\varsigma_j^2$ is the MSE of string $j$ that all CAVs are benign; $n$ is the number of strings. If the MSE is lower than $\tau$, the string is considered as a benign set that doesn't include attacked vehicle(s). Therefore, our goal is to identify the largest benign set of CAVs' distance.

To reduce computational loads of the defense strategy, we propose a step-wise deletion greedy algorithm (as shown in Algorithm 3). This greedy algorithm starts with the initial location set including all CAVs of the string. In subsequent time steps, the algorithm will keep verifying if the MSE of current set of CAVs' locations is lower than the threshold. If yes, the set is confirmed for further control. Otherwise, the algorithm computes MSEs of all possible sets and chooses the subset with the least MSE as the input to the next time step. This algorithm continues until it finds the set that meets the threshold condition.

## SIMULATION STUDY AND RESULTS

In this section, we describe the simulation settings and parameters, evaluate the system performance, including mobility, safety, and environment under different scenarios (e.g., with and without cyber-attacks as well as the defense strategy), and analyze the simulation results. We choose one traffic demand as our benchmark and evaluate the improvement or deterioration compared to the baseline (no CAVs and thus no cyber-attacks).

We set up the simulation environment with VENTOS, using the network shown in Figure 3(a). The only transmission noise is "thermal noise," which is set to be -90dBm. To make the simulation more realistic, we add noise to the vehicles' positions in the simulation based on a random walk model. The transmission power of 20 mW and the data rate of 6 Mbps are chosen as the default values in VENTOS. There are two lanes on the highway segment and one lane on the on-ramp. Only CAVs on the rightmost lane of the highway and the on-ramp (i.e., involved in merging maneuvers) would play a major role in this simulation controlled by the merging algorithm. With VENTOS, we can also visualize the BSM transmission, as shown in Figure 3(b). As aforementioned, we assume the RSSI cannot be compromised. The vehicles on the highway can change their lane freely based on the default lane-changing algorithm in SUMO. The attacks start only when the target CAV enters the attacker's effective area, and stop when it exits the area. For those CAVs traveling along the mainline, we assume that attacks can be deployed immediately once the target CAVs are traveling on the rightmost lane and continuously take effects even they change to the left lane afterwards, as long as they are still within the attacker's range. In the simulation study, we set the traffic demand ratio between highway and on-ramp to be 3:1, and the roadway capacity to be 2000 passenger car units per hour per lane (pcu/hr/ln). With different volume-to-capacity (V2C) ratios, we schedule each vehicle's departure time using a Poisson distribution.

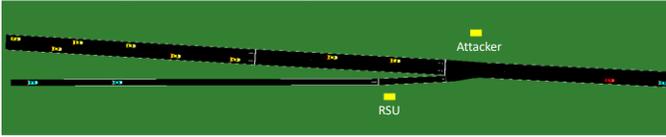
(a) Simulation network of the highway and the ramp

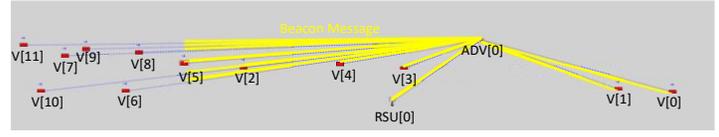
(b) BSM transmission visualization

**Figure 3 VENTOS Simulation network and visualization.**

We evaluate the performance of mobility with four CAV penetration rates (i.e., 0%, 20%, 50% and 100%), three V2C ratios (i.e., 0.3, 0.6 and 0.9), and four attack ratio (i.e., 0, 0.1, 0.25, and 0.5) which represent the percentages of attacked CAVs with respect to the entire CAV population. Since evaluating nonzero attack ratio in a non-CAV scenario is meaningless, we focus on 39 different cases in total. The simulation time for each run is set to be 20 minutes. The mobility performance is measured by the network efficiency,

$$Q = \frac{VMT}{VHT} \quad (11.)$$

where VMT is the total vehicle-miles traveled in the network, and VHT is the total vehicle-hours traveled in the network accordingly. Efficiencies of three typical cases are shown in Figure 4. Figure 4(a) shows the efficiency as a mapping of both attack ratio and V2C ratio under 100% CAV penetration rate. It can be observed that efficiencies decrease as the attack ratio increases. From both Figure 4(a) and Figure 4(b), it can be observed that as the V2C ratio decreases, efficiencies get increased. In the case without attack, we note the positive correlation between penetration rate and efficiency. By observing Figure 4(c) where attack ratio is 0.5, we notice that the results are quite different from Figure 4(b). Because of the attack, the system performance would vary under different V2C ratios. When the traffic is light, benefits from the increasing penetration rate of CAVs could not offset the negative impacts due to cyber-attacks. Therefore, the system efficiency gets lowered as the penetration rate increases. When the traffic gets more congested, the system efficiency may reach its peak at the penetration rate level of 50%.

**Table 1 Mobility performance of traffic flow of selected simulation cases**

| Case Index | Penetration Rate | Attack Ratio | V2C Ratio | VMT (mile) | VHT (hour) | Efficiency (mph) |
|---|---|---|---|---|---|---|
| 1 | 100 | 0.5 | 0.9 | 26 | 2.4 | 10.9 |
| 2 | 50 | 0.5 | 0.9 | 68.7 | 4.5 | 15.4 |
| 3 | 20 | 0.5 | 0.9 | 89.1 | 6.8 | 13.1 |
| 4 | 100 | 0.25 | 0.9 | 38.9 | 2.6 | 14.9 |
| 5 | 50 | 0.25 | 0.9 | 68.8 | 4.5 | 15.2 |
| 6 | 20 | 0.25 | 0.9 | 90.9 | 7.2 | 12.7 |
| 10 | 100 | 0 | 0.9 | 97.7 | 4 | 24.4 |
| 27 | 100 | 0.5 | 0.3 | 31.2 | 2.2 | 14.2 |
| 28 | 50 | 0.5 | 0.3 | 60.2 | 3.3 | 18.5 |
| 29 | 20 | 0.5 | 0.3 | 70.7 | 2.9 | 24.8 |
| 31 | 50 | 0.25 | 0.3 | 63.2 | 2.9 | 22.1 |
| 34 | 50 | 0.10 | 0.3 | 64.1 | 2.7 | 23.5 |
| 37 | 50 | 0 | 0.3 | 67 | 2.5 | 26.9 |

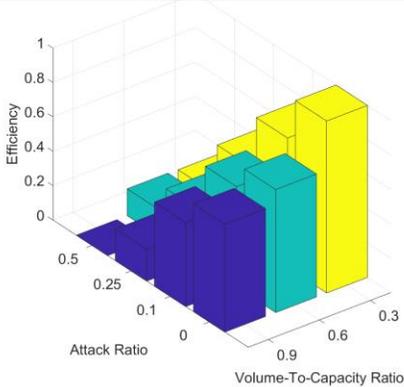
(a) 100% Penetration Rate

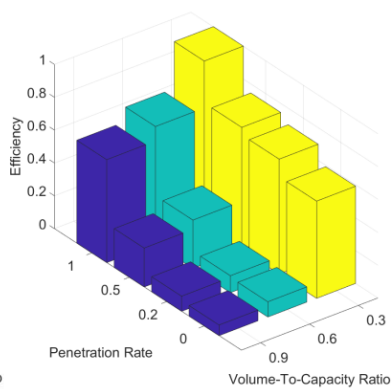
(b) Zero Attack Ratio

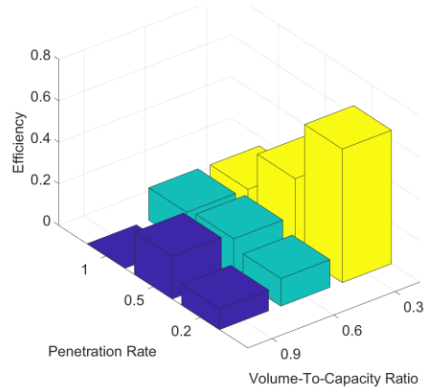
(c) 0.5 Attack Ratio

**Figure 4 Bar charts of efficiencies in different cases.**

Table 1 summarizes the overall mobility performance of selected cases. As expected, in a non-attack scenario with a fixed V2C ratio, higher penetration rates mean higher mobility efficiencies. However, this observation does not hold under attacks. As shown in Cases 27, 28, and 29 (i.e., with 0.5 attack ratio), efficiencies decrease significantly with increasing CAV penetration rates. When the attack ratio is 0.25 (e.g., Cases 4, 5, and 6), the difference in efficiency is not significant, but VMT drops rapidly when the penetration rate increases. This indicates that the traffic gets more congested, and spawning of vehicles in simulation is even blocked. Under the same penetration rate, as the V2C ratio is reduced, deploying more attacks can reduce efficiency (see Cases 28, 31, 34, and 37). With the same attack ratio (e.g., 0.25 or 0.5) and a high V2C ratio (e.g., 0.9), efficiencies show slight fluctuations, as indicated in Cases 1, 2, 3, and 4. A hypothesis is that when the traffic volume reaches the road capacity, vehicles are moving slowly on the entire approach. In these cases, increasing the attack ratio introduces more chaos into the network, reducing the traffic throughput at the bottleneck (i.e., the ramp merging area), as observed from the VMT values. In particular, the most congested case is Case 1, whose efficiency is only 10.9 mph, 55.19% worse than its counterpart – Case 10 (with the same V2C ratio and penetration rate but no attacks).

On the other hand, we quantify the safety risks due to cyber-attacks in terms of driving volatility which represents the stability of vehicle movement. In this paper, "Standard Deviation" and "Mean Absolute Deviation" measurements [42] are used for computing the driving volatility, which are defined below,

$$S_{dev} = \sqrt{\frac{1}{n-1}\sum_{i=1}^{n}(x_i - \bar{x})^2} \quad (12.)$$

$$D_{mean} = \frac{1}{n}\sum_{i=1}^{n}|x_i - \bar{x}| \quad (13.)$$

where $n$ is the total number of observations, $x_i$ represents the $i^{th}$ of observations, $\bar{x}$ is the mean of observations and $S_{dev}$ is the standard deviation. Larger deviation means higher driving volatility.

**Table 2: Safety performance of traffic flow of selected simulation cases**

| Case Index | Penetration Rate (%) | Attack Ratio | V2C Ratio | Velocity Mean Absolute Deviation (m/s) | Velocity Standard Deviation (m/s) | Accel Mean Absolute Deviation (m/s^2) | Accel Standard Deviation (m/s^2) |
|---|---|---|---|---|---|---|---|
| 28 | 50 | 0.5 | 0.3 | 5.81 | 5.14 | 3.05 | 3.76 |
| 31 | 50 | 0.25 | 0.3 | 5.15 | 3.96 | 3.47 | 4.07 |
| 33 | 100 | 0.1 | 0.3 | 5.86 | 4.37 | 2.98 | 3.82 |
| 34 | 50 | 0.1 | 0.3 | 2.98 | 2.56 | 3.53 | 4.12 |
| 35 | 20 | 0.1 | 0.3 | 2.68 | 1.84 | 4 | 4.46 |
| 36 | 100 | 0 | 0.3 | 2.19 | 1.05 | 2.89 | 3.72 |
| 37 | 50 | 0 | 0.3 | 2.38 | 1.38 | 3.78 | 4.3 |
| 38 | 20 | 0 | 0.3 | 2.69 | 1.46 | 3.99 | 4.45 |
| 39 | 0 | 0 | 0.3 | 2.96 | 1.93 | 4.05 | 4.47 |

As shown in Table 2, deviations of velocity and acceleration decrease when penetration rates rise under the scenarios of non-attack and fixed traffic volumes. In particular, when the penetration rate is 50% (i.e., Case 28, 31, 34, and 37), the deviation of acceleration reduces while the deviation of velocity increases, as the attack ratio grows. Please note that high velocity deviation means instability in the traffic.

We further analyze the energy consumption and pollutant emissions with the U.S. Environmental Protection Agency's MOtor Vehicle Emission Simulator (MOVES) [43]. The selected results are summarized in Table 3. From this table, it can be observed that with the increase of the CAV penetration rate, the fuel consumption and CO2 emissions decrease, which matches the conclusion by Wang et al.[5]. According to Cases 28, 31, 34, and 37, we can conclude that the fuel consumption, CO2 and NOx emissions are positively correlated with the attack ratio

**Table 3: Energy performance of traffic flow of selected simulation cases**

| Case Index | Penetration Rate (%) | Attack Ratio | V2C Ratio | Fuel (g/mi) | CO (g/mi) | HC (g/mi) | NOx (g/mi) | PM2.5 (g/mi) | CO2 (g/mi) |
|---|---|---|---|---|---|---|---|---|---|
| 28 | 50 | 0.5 | 0.3 | 187.6 | 1.6515 | 0.0137 | 0.0729 | 0.0036 | 598.9 |
| 31 | 50 | 0.25 | 0.3 | 178.3 | 1.6376 | 0.0134 | 0.0712 | 0.0034 | 569.5 |
| 34 | 50 | 0.1 | 0.3 | 173.3 | 1.7115 | 0.0137 | 0.0700 | 0.0035 | 553.5 |
| 37 | 50 | 0 | 0.3 | 166.0 | 1.5427 | 0.0126 | 0.0672 | 0.0032 | 530.2 |

To investigate the system performance under the other attack strategy and defense algorithm, we select the benchmark scenario with 900 pcu/hr/ln on the highway and 300 pcu/hr/ln on the on-ramp under 0.3 V2C ratio. The baseline case is set with the same demand, 0% penetration rate, and 0 attack ratio, whose time-space diagram is shown in Figure 5(a). The total VMT and VHT of the baseline case are 67.1 miles and 2.8 hours, respectively, and the network efficiency is 23.7 mph. Figure 5(b) presents the time-space diagram for the scenario with 50% CAV penetration rate. Compared to the baseline, the network efficiency increases by 13.9% (to 26.9 mph). As shown in the zoom-in parts of the time-space diagrams at the ramp merging area, application of the proposed merging strategy can significantly help mitigate upstream shockwaves and smooth vehicle trajectories.

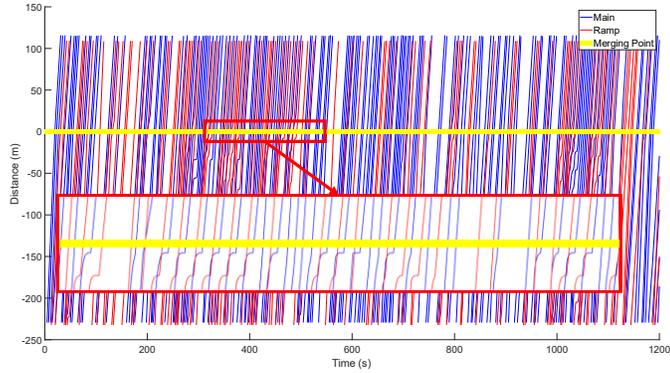 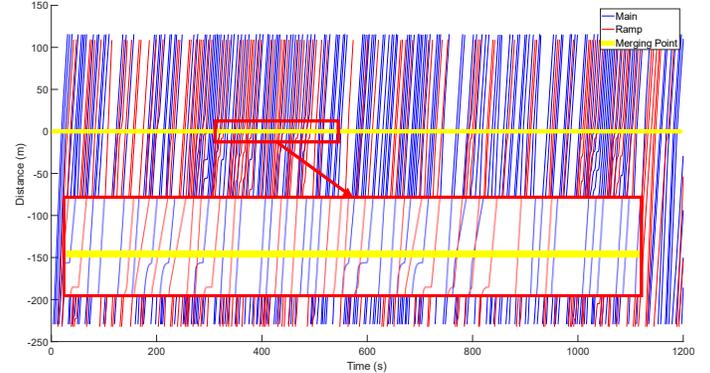

(a) Baseline  (a) 50% penetration rate

**Figure 5 Time-space diagrams of two cases with different penetration rates with 0 attack ratio and 30% V2C ratio**

Next, we enable the attacker's behaviors with a 0.5 attack ratio. Under the first attack strategy, VMT is decreased to 60.2 miles, VHT is reduced to 3.3 hours, and the network efficiency is only 18.5 mph. With the same attack ratio, we deploy the second attack strategy, i.e., accumulative position drift spoofing and the traffic gets more congested. The efficiency is 17.9 mph, and VMT and VHT are 48.6 miles and 2.7 hours, respectively. The time-space diagram of the second attack strategy is shown in Figure 6. In Figure 6, it can be observed from the zoom-in area that the shockwaves created by attacks occur much more frequently and last much longer time compared to the case without attacks. We also deploy the proposed defense algorithm to detect and filter out the malicious information created by attacks. Due to the introduction of defense algorithm, the network efficiencies under the first attack strategy and second attack strategy become 26.2 mph and 25 mph, respectively, both of which are much better than the associated cases without defense and almost close to the cases without attacks. Figure 7 shows the time-space diagram after implementing the defense strategy under the second attack strategy. As represented by the smoother trajectories and fewer shockwaves in the figure, the defense strategy can help alleviate the congestion caused by attacks. Table 4 summarizes key statistics of representative scenarios.

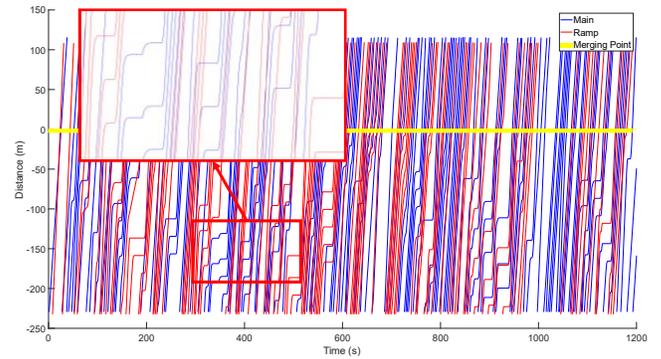

**Figure 6 Time-space diagrams under the accumulative position drift spoofing attacks with 0.5 attack ratio.**

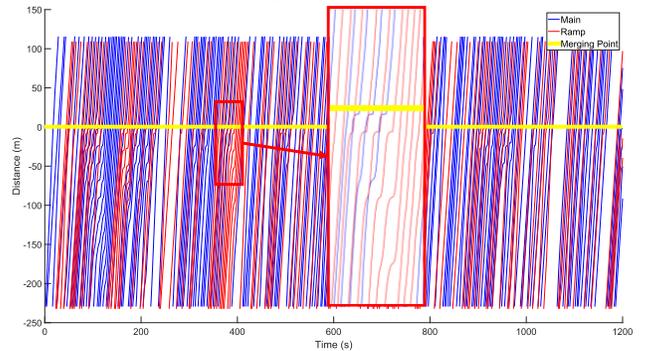

**Figure 7 Time-space diagrams under the second type of attacks with defense strategy**

**Table 4: Mobility performance of traffic flow of each simulation cases**

|  | VMT (mile) | VHT (hour) | Efficiency (mph) |
|---|---|---|---|
| **Baseline (0% Penetration Rate, 0 Attack Ratio, 30% V2C Ratio)** | 67.1 | 2.8 | 23.7 |
| **Cooperative merging system (50% Penetration Rate)** | 67 | 2.5 | 26.9 |
| **Emergency Stop Spoofing (0.5 Attack Ratio)** | 60.2 | 3.3 | 18.5 |
| **Accumulative Position Drift Spoofing (0.5 Attack Ratio)** | 48.6 | 2.7 | 17.9 |
| **Emergency Stop Spoofing with Defense algorithm** | 64.6 | 2.5 | 26.2 |
| **Accumulative Position Drift Spoofing with Defense algorithm** | 73 | 2.9 | 25 |

## CONCLUSIONS AND FUTURE WORK

This study reveals the cybersecurity risks of a typical CAV application, i.e., cooperative highway on-ramp merging in a mixed traffic environment. Two non-trivial cyber-attack strategies, i.e., emergency stop spoofing and accumulative position drift spoofing, have been proposed and deployed in VENTOS. Simulation results of mobility, safety, and environmental sustainability for 39 cases with different CAV penetration rates, V2C ratios, and cyber-attack ratios have been compared and analyzed. In the worst case, up to 55.19% decrease in network efficiency is observed. Unlike the scenarios without cyber-attacks, cases with higher CAV penetration rates are more susceptible to the presence of attacks, leading to significant system performance degradation. To address this issue, an MMSE-based defense algorithm is proposed and deployed in this study. Simulation results indicate that the proposed defense algorithm can well improve the cyber-attack resilience of the system. It can recover most benefits from the cooperative merging system under two attack strategies and even perform better than non-CAV scenarios.

As one of the future steps, risks of other types of cyber-attacks such as jamming (DOS or DDOS), black-hole attacks, and Sybil attacks will be devised and evaluated for cooperative highway on-ramp merging scenarios. The cybersecurity performance of other CAV applications will be also investigated, and the respective defense strategies should be designed to further improve the attack-resilience performance of target CAV systems.

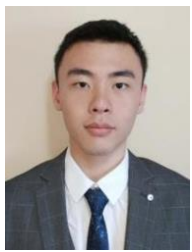
**Xuanpeng Zhao** received the B.E. degree in electrical engineering from Shanghai Maritime University in 2019 and the M.S. degree in electrical engineering from University of California at Riverside. He is currently a Ph.D. student in electrical engineering at University of California at Riverside. His research focuses on embedded system, computer vision, and connected and automated vehicle technology.

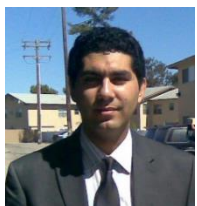
**Ahmed Abdo** obtained his BS in electrical engineering at Menofia University, Egypt, and MSEE at California State University, Los Angeles, USA. He is currently a PhD student in computer engineering at UCR, USA. His research interests center around security in automated and autonomous systems such as connected and self driven vehicles.

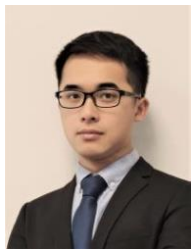
**Xishun Liao** (S'19) received the B.E. degree in mechanical engineering and automation from Beijing University of Posts and Telecommunications in 2016, and the M.Eng. degree in mechanical engineering from University of Maryland, College Park in 2018. He is currently a Ph.D. student in electrical and computer engineering at University of California at Riverside. His research focuses on connected and automated vehicle technology.

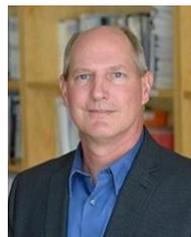
**Matthew J. Barth** (M'90-SM'00-F'14) received the M.S. and Ph.D. degree in electrical and computer engineering from the University of California at Santa Barbara, in 1985 and 1990, respectively. He is currently the Yeager Families Professor with the College of Engineering, University of California at Riverside, USA. He is also serving as the Director for the Center for Environmental Research and Technology. His current research interests include ITS and the environment, transportation/emissions modeling, vehicle activity analysis, advanced navigation techniques, electric vehicle technology, and advanced sensing and control. Dr. Barth has been serving as a Senior Editor for both the Transactions of ITS and the Transactions on Intelligent Vehicles. He served as the IEEE ITSS President for 2014 and 2015 and is currently the IEEE ITSS Vice President for Finance.

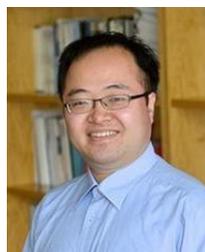
**Guoyuan Wu** (M'09-SM'15) received his Ph.D. degree in mechanical engineering from the University of California, Berkeley in 2010. Currently, he holds an Associate Researcher and an Associate Adjunct Professor position at Bourns College of Engineering – Center for Environmental Research & Technology (CE–CERT) and Department of Electrical & Computer Engineering in the University of California at Riverside. development and evaluation of sustainable and intelligent transportation system (SITS) technologies, including connected and automated transportation systems (CATS), shared mobility, transportation electrification, optimization and control of vehicles, traffic simulation, and emissions measurement and modeling. Dr. Wu serves as an Associate Editor for IEEE Transactions on Intelligent Transportation Systems, SAE International Journal of Connected and Automated Vehicles, and IEEE Open Journal of ITS. He is also a member of the Vehicle-Highway Automation Standing Committee (ACP30) of the Transportation Research Board (TRB), and a recipient of Vincent Bendix Automotive Electronics Engineering Award.